\documentclass{chjaa}

\usepackage{graphicx,times,natbib,url}

\def\ltsima{$\; \buildrel < \over \sim \;$}
\def\lsim{\lower.5ex\hbox{\ltsima}}

\bibpunct{(}{)}{;}{a}{}{,}

\begin{document}

\title{The afterglow onset for GRB\,060418 and GRB\,060607A}

\volnopage{Vol.0 (200x) No.0, 000--000}      
\setcounter{page}{1}

\author{S.~Covino\inst{1}\mailto{}
\and S.D.~Vergani\inst{2,3}
\and D.~Malesani\inst{4,5}
\and E.~Molinari\inst{1}
\and P.~D'Avanzo\inst{6,1}
\and G.~Chincarini\inst{7,1}
\and F.M.~Zerbi\inst{1}
\and L.A.~Antonelli\inst{8,9}
\and P.~Conconi\inst{1}
\and V.~Testa\inst{8}
\and G.~Tosti\inst{10}
\and F.~Vitali\inst{8}
\and F.~D'Alessio\inst{8}
\and G.~Malaspina\inst{1}
\and L.~Nicastro\inst{11}
\and E.~Palazzi\inst{11}
\and D.~Guetta\inst{8}
\and S.~Campana\inst{1}
\and P.~Goldoni\inst{12,13}
\and N.~Masetti\inst{11}
\and E.J.A.~Meurs\inst{2}
\and A.~Monfardini\inst{14}
\and L.~Norci\inst{3}
\and E.~Pian\inst{15}
\and S.~Piranomonte\inst{8}
\and D.~Rizzuto\inst{1,7}
\and M.~Stefanon\inst{16}
\and L.~Stella\inst{8}
\and G.~Tagliaferri\inst{1}
\and P.A.~Ward\inst{2}
\and G.~Ihle\inst{16}
\and L.~Gonzalez\inst{16}
\and A.~Pizarro\inst{16}
\and P.~Sinclair\inst{16}
\and J.~Valenzuela\inst{16}
}

\institute{INAF - Osservatorio Astronomico di Brera, via E. Bianchi 46, I-23807 Merate (LC), Italy. \\
\email{stefano.covino@brera.inaf.it}
\and Dunsink Observatory - DIAS, Dunsink lane, Dublin 15, Ireland.\\
\and School of Physical Sciences and NCPST, Dublin City University - Dublin 9, Ireland.\\
\and International School for Advanced Studies (SISSA/ISAS), via Beirut 2-4, I-34014 Trieste, Italy.\\
\and Dark Cosmology Centre, Niels Bohr Institute, University of Copenhagen, Juliane Maries vej 30, DK--2100 K\o{}benhavn, Denmark.\\
\and Dipartimento di Fisica e Matematica, Universit\`a dell'Insubria, via Valleggio 11, I-22100 Como, Italy.\\
\and Universit\`a degli Studi di Milano Bicocca, piazza delle Scienze 3, I-20126 Milano, Italy.\\
\and INAF - Osservatorio Astronomico di Roma, via di Frascati 33, I-00040 Monteporzio Catone (Roma), Italy.\\
\and ASI Science Data Center, via G. Galilei, I-00044 Frascati (Roma), Italy \\
\and Dipartimento di Fisica e Osservatorio Astronomico, Universit\`a di Perugia, via A. Pascoli, I-06123 Perugia, Italy.\\
\and INAF-IASF di Bologna, via P. Gobetti 101, I-40129 Bologna, Italy.\\
\and APC, Laboratoire Astroparticule et Cosmologie, UMR 7164, 11 Place Marcelin Berthelot, F-75231 Paris Cedex 05, France.\\
\and CEA Saclay, DSM/DAPNIA/Service d'Astrophysique, F-91191, Gif-s\^ur-Yvette, France.\\
\and CNRS, Institut Néel, 25 rue des Martyrs, F-38042 Grenoble, France.\\
\and INAF - Osservatorio Astronomico di Trieste, via G.B. Tiepolo 11, I-34143 Trieste, Italy.\\
\and European Southern Observatory, Alonso de C\'ordova 3107, Vitacura, Casilla 19001, Santiago 19, Chile.\\
}

\date{Received; accepted}

\abstract{
Gamma-ray burst are thought to be produced by highly relativistic outflows. Although
upper and lower limits for the outflow initial Lorentz factor $\Gamma_0$ are available, observational
efforts to derive a direct determination of $\Gamma_0$ have so far failed or provided ambiguous 
results. As a matter of fact, the shape of the early-time afterglow light curve is strongly sensitive on $\Gamma_0$ 
which determines the time of the afterglow peak, i.e. when the outflow and the shocked circumburst
material share a comparable amount of energy.
We now comment early-time observations of the near-infrared afterglows of GRB\,060418 and GRB\,060607A
performed by the REM robotic telescope.
For both events, the afterglow peak was singled out and allowed us to determine the initial fireball Lorentz,  
$\Gamma_0\sim400$.
}

\titlerunning{The onset of the afterglow}
\authorrunning{Covino et al.}

\maketitle

\section{Introduction}         

The early stages of gamma-ray burst (GRB) afterglow light curves display a rich
variety of phenomena at all wavelengths and contain significant information
which may allow determining the physical properties of the emitting fireball.
The launch of the \textit{Swift} satellite \citep{Neil04}, combined with the
development of fast-slewing ground-based telescopes, has hugely improved the
sampling of early GRB afterglow light curves.

Since many processes work in the early afterglow, it is often difficult to
model them well enough to be able to determine the fireball characteristics.
The simplest case is a light curve shaped by the forward shock only. 
This case is particularly
interesting because, while the late-time light curve is independent of the
initial conditions (the so-called self-similar solution), the time at which the afterglow
peaks depends on the original fireball Lorentz factor $\Gamma$, thus allowing a
direct measurement of this fundamental parameter \citep{SP99}. The short
variability timescales, coupled with the nonthermal GRB spectra, indeed imply
that the sources emitting GRBs have a highly relativistic motion
\citep{Ruderman75,Fenimore93,Piran00,Lith01}, to avoid suppression of the
high-energy photons due to pair production. This argument, however, can only
set a lower limit to the fireball Lorentz factor. Late-time measurements (weeks
to months after the GRB) have shown $\Gamma \sim \mbox{a few}$
\citep{Frail97,Taylor05}, but a direct measure of the initial value (when
$\Gamma$ is expected to be $\sim 100$ or more) is still lacking.

We present here the NIR early light curves of the GRB\,060418 and GRB\,060607A
afterglows observed with the REM robotic
telescope\footnote{\url{http://www.rem.inaf.it}} \citep{Zerb01,Chinc03} located
in La Silla (Chile). These light curves show the onset of the afterglow and its
decay at NIR wavelengths as simply predicted by the fireball forward shock
model, without the presence of flares or other peculiar features. A detailed discussion
of these data has also been reported by \citet{Mol07} and \citet{Mal07}.

\begin{figure}
\centering
\includegraphics[width=11cm]{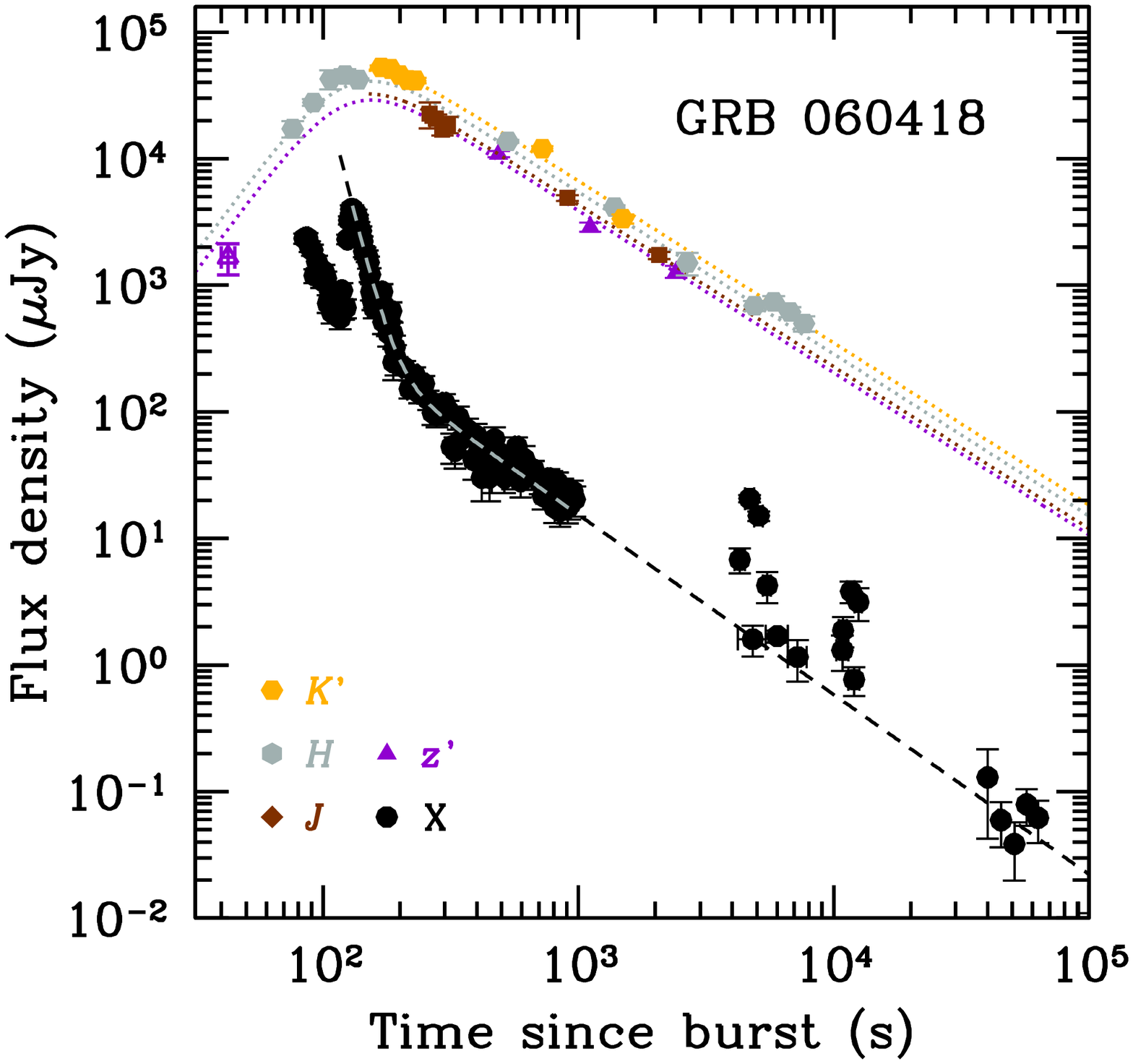} \\
\centering\includegraphics[width=11cm]{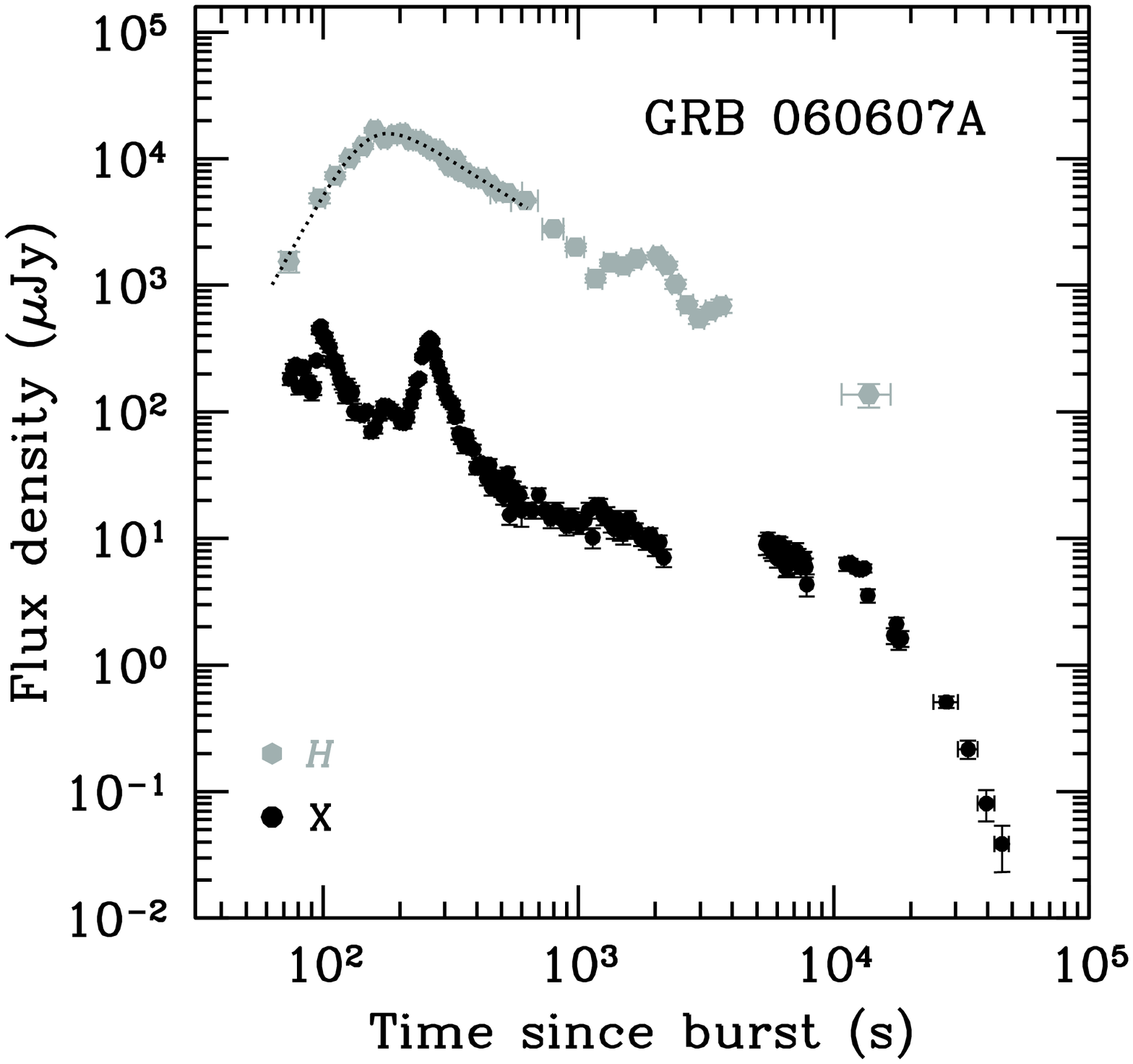}
\caption{NIR and X-ray light curves of GRB\,060418 (upper panel) and 
GRB\,060607A (lower panel). The dotted lines show the
models of the NIR data using the smoothly broken power law (see Sect.
\ref{modelling}).
For GRB\,060418 the dashed line shows the best-fit to the X-ray data.\label{fig:lcs}}
\end{figure}

\section{Observations}

GRB\,060418 and GRB\,060607A were detected by \textit{Swift} at 03:06:08\,UT
\citep{Falc06a} and 05:12:13\,UT \citep{Ziaee06}, respectively. The BAT light
curve of the former ($T_{90} = 52 \pm 1$~s) showed three overlapping peaks
\citep{Cumm06}. For the latter, the light curve is dominated by a double-peaked
structure with a duration $T_{90} = 100 \pm 5$\,s \citep{Tuel06}. The
\textit{Swift} XRT started observing the fields 78 and 65~s after the trigger,
respectively.  UVOT promptly detected bright optical
counterparts for both events. The redshift is $z=1.489$ for GRB\,060418
\citep{Dupr06,Vrees06} and $z=3.082$ for GRB\,060607A \citep{Led06}.

The REM telescope reacted promptly to both GCN alerts and began observing the
field of GRB\,060418 64\,s after the burst (39\,s after the reception of the
alert) and the field of GRB\,060607A 59\,s after the burst (41\,s after the
reception of the alert). For both targets a bright NIR source was identified
\citep{Cov06a,Cov06b}.

\section{Results and Discussion}

\subsection{Light Curve Modelling}
\label{modelling}

\begin{table}
\caption{Best Fit Values of the Light Curves of the First Hour of Observations
for GRB\,060418 and the First 1000\,s for GRB\,060607A (1$\sigma$ Errors),
Using a Smoothly Broken Power-Law. Fit Parameters are Defined in \citet{Mol07}. 
The Relatively Large $\chi^2$
of the Fit Results from Small-Scale Irregularities Present Throughout the Light
Curve (see Fig. \ref{fig:lcs}).}
\centering
\begin{tabular*}{\columnwidth}{@{\extracolsep{\fill}}l@{\hfill}c@{\hfill}c@{\hfill}c@{\hfill}c@{\hfill}c@{\hfill}c}
\hline
GRB     & $t_{\rm peak}$ (s) & $t_{\rm b}$   (s)     & $\alpha_{\rm r}$      & $\alpha_{\rm d}$       & $\kappa$            & $\chi^2/{\rm d.o.f.}$ \\  \hline
060418  & $153_{-10}^{+10}$ & $127_{-21}^{+18}$  & $-2.7_{-1.7}^{+1.0}$  & $1.28_{-0.05}^{+0.05}$ & $1.0_{-0.4}^{+0.4}$ & $33.3/16$             \\
060607A & $180_{-6}^{+5}$   & $153_{-12}^{+12}$  & $-3.6_{-1.1}^{+0.8}$  & $1.27_{-0.11}^{+0.16}$ & $1.3_{-1.1}^{+0.9}$ & $28.5/19$             \\ 
\hline
\end{tabular*}

\label{tab:fit}
\end{table}

Figure~\ref{fig:lcs} show
the NIR and X-ray light curves of the two afterglows. The X-ray data have
been taken with the \textit{Swift} XRT. The NIR light curves of the two events show a
remarkable similarity. Both present an initial sharp rise, peaking at
100--200~s after the burst. The NIR flux of GRB\,060418 decays afterwards as a
regular power law. The NIR light curve of GRB\,060607A shows a similar, smooth
behaviour up to $\sim 1000$~s after the trigger, followed by a rebrightening
lasting $\sim 2000$~s. 

To quantitatively evaluate the peak time, we fitted the NIR light curves using
a smoothly broken power-law \citep{Beu99}. 
We obtain for GRB\,060418 and GRB\,060607A peak times of
$153 \pm 10$ and $180 \pm 6$~s, respectively. 
The complete set of fit results is reported
in Table~\ref{tab:fit}. 

As for many other GRBs observed by \textit{Swift}, the early X-ray light curves
of both events show several, intense flares superimposed on the power-law decay
\citep{Chinca07}. In particular, for GRB\,060418 a bright flare was active
between $\sim 115$ and 185~s. Excluding flaring times, the decay is then
described by a power law with decay slope $\alpha_{\rm X} = 1.42 \pm 0.03$. The
X-ray light curve of GRB\,060607A is more complex and presents two large flares
within the first 400~s. After that the flux density decreases following a
shallow power law (with small-scale variability), until steepening sharply at
$t \sim 10^4$~s.

By comparing the X-ray and NIR light curves of both bursts, it is apparent that
the flaring activity, if any, is much weaker at NIR frequencies. It is thus
likely that the afterglow peak, visible in the NIR, is hidden in the X-ray
region.

\subsection{Determination of the Lorentz Factor $\Gamma_0$}
\label{sec:lor}

Our spectral and temporal analysis agrees with the interpretation of the NIR
afterglow light curves as corresponding to the afterglow onset, as predicted by
the fireball forward shock model \citep{SP99,Meszaros}. According to the fits,
the light curves of the two afterglows peak at a time $t_{\rm peak} > T_{90}$
as expected in the impulsive regime outflow (`thin shell' case). In this
scenario the quantity $t_{\rm peak}/(1+z)$ corresponds to the deceleration
timescale $t_{\rm dec} \sim R_{\rm dec}/(2c\Gamma_{\rm dec}^2)$, where $R_{\rm
dec}$ is the deceleration radius, $c$ is the speed of light and $\Gamma_{\rm
dec}$ is the fireball Lorentz factor at $t_{\rm dec}$. It is therefore possible
to estimate $\Gamma_{\rm dec}$ \citep{SP99}, which is expected to be half of
the initial value $\Gamma_0$ \citep{PK00,Meszaros}. For a homogeneous
surrounding medium with particle density $n$, we have
\begin{equation}
\Gamma(t_{\rm peak}) = \left[ \frac{3E_{\gamma}(1+z)^3}{32\pi n m_{\rm p} c^5 
\eta t_{\rm peak}^3} \right]^{1/8}\approx 160 \left[\frac{E_{\gamma,53}(1+z)^3}{\eta_{0.2}n_0 {t}_{\rm peak,2}^3}\right]^{1/8},
\label{Gamma}
\end{equation}
where $E_\gamma = 10^{53} E_{\gamma,53}$~erg is the isotropic-equivalent 
energy released by the GRB in gamma rays, $n = n_0$~cm$^{-3}$, $t_{\rm peak,2}
= t_{\rm peak}/(100~{\rm s})$, $\eta = 0.2\,\eta_{0.2}$ is the radiative
efficiency and $m_{\rm p}$ is the proton mass. We use
$E_{\gamma}=9\times10^{52}$\,erg for GRB\,060418 \citep{Golen06} and
$E_{\gamma}\sim1.1\times10^{53}$\,erg for GRB\,060607A \citep{Tuel06}.
Substituting the measured quantities and normalising to the typical values $n =
1$~cm$^{-3}$ and $\eta = 0.2$ \citep{Bloom03}, we infer for both bursts 
$\Gamma_0 \approx 400\,(\eta_{0.2}n_0)^{-1/8}$.

\subsubsection{How Model Dependent is the Derived Lorentz Factor?}

In the context of the so-called standard afterglow model, the Lorentz $\Gamma$ factor determined 
in Sect.\,\ref{sec:lor} is only very weakly dependent on the unknown parameters $n$ and $\eta$. 
Therefore, the determination of $\Gamma_0$ is robust. In principle, the only important factor is the
hydrodynamical interaction of a relativistic outflow with the circumburst medium. Independently of 
the emission process or of the interaction physics, if the material collected by the outflow
is able to radiate away the acquired energy, the phenomenon should be qualitatively the same, 
and again the Lorentz $\Gamma$ factor estimate would be reliable. Of course, in this hypothetical 
case, lacking of a well developed theoretical framework we could not check
the self-consistency of the proposed scenario studying the slopes of the rising (and decaying) 
phase.

\subsection{The Reverse Shock}

For both bursts, we could not detect any reverse shock emission. The lack of
such flashes has already been noticed previously in a set of \textit{Swift}
bursts with prompt UVOT observations \citep{Rom06}. Among the many possible
mechanisms to explain the lack of this component, strong suppression (or even
total lack) of reverse shock emission is naturally expected if the outflow is
Poynting-flux dominated \citep{Fan04,ZhangKob05}. For GRB\,060418, \citet{Mun07}
derived an 8\% upper limit for the polarization in the optical band roughly three minutes from 
the burst, i.e. one minute after the afterglow onset. In case the GRB prompt is driven 
by magnetic energy a high polarization degree is expected \citep{GrKo03,Laz04,Sag04}. However,
for photons emitted by material shocked by the forward shock the polarization degree would
depend on the magnetic energy transfer from the blastwave to the shocked medium, that is at 
present poorly known \citep[see][and references therein]{Cov07}. As a matter of fact, \cite{Jin07}
showed that for GRB\,060418 and GRB\,060607A the reverse shock emission predicted by the standard
afterglow model might be too weak to be detected.

\section{Conclusions}

The REM discovery of the afterglow onset has demonstrated once again the
richness and variety of physical processes occurring in the early afterglow
stages. The very fast response observations presented here provide crucial
information on the GRB fireball parameters, most importantly its initial
Lorentz factor. This is the first time that $\Gamma(t_{\rm peak})$ is directly
measured from the observations of a GRB. The measured $\Gamma_0$ value is well
within the range ($50 \la \Gamma_{0} \la 1000$) envisaged by the standard
fireball model \citep{Piran00,Guetta01,Alicia02,Meszaros}. It is also in
agreement with existing measured lower limits \citep{Lith01,zhang06}. 

Using $\Gamma_0=400$ we can also derive other fundamental quantities
characterising the fireball of the two bursts. In particular, the
isotropic-equivalent baryonic load of the fireball is $M_{\rm fb} = E/(\Gamma_0
c^2) \approx 7\times 10^{-4}\,M_\odot$, and the deceleration radius is $R_{\rm
dec} \approx 2 c t_{\rm peak} [\Gamma(t_{\rm peak})]^2/(1+z) \approx
10^{17}$~cm. This is much larger than the scale of $\sim 10^{15}$~cm where the
internal shocks are believed to power the prompt emission
\citep{MesRees97,Rees94}, thus providing further evidence for a different
origin of the prompt and afterglow stages of the GRB.


\label{lastpage}


\begin{thebibliography}{}
\bibitem[Beuermann et al.(1999)]{Beu99} Beuermann, K., Hessman, F. V., Reinsch, K., et al. 1999, A\&A, 352, L26 
\bibitem[Bloom et al.(2003)]{Bloom03} Bloom, J.~S., Frail, D.~A., \& Kulkarni, S.~R. 2003, ApJ, 594, 674
\bibitem[Chincarini et al.(2003)]{Chinc03} Chincarini, G., Zerbi, F. M., Antonelli, A., et al. 2003, The Messenger, 113, 40
\bibitem[Chincarini et al.(2007)]{Chinca07} Chincarini, G., Moretti, A., Romano, P., et al. 2007, \apj, submitted (astro-ph/0702371) 
\bibitem[Covino(2007)]{Cov07} Covino, S. 2007, Science, 315, 1798
\bibitem[Covino et al.(2006a)]{Cov06a} Covino, S., Antonelli, L. A., Vergani, S. D., et al. 2006a, GCN\,4967
\bibitem[Covino et al.(2006b)]{Cov06b} Covino, S., Distefano, E., Molinari, E., et al. 2006b, GCN\,5234
\bibitem[Cummings et al.(2006)]{Cumm06} Cummings, J., Barbier, L., Barthelmy, S. et al. 2006, GCN\,4975
\bibitem[Dupree et al.(2006)]{Dupr06} Dupree, A. K., Falco, E., Prochaska, J. X., et al. 2006, GCN\,4969
\bibitem[Falcone et al.(2006)]{Falc06a} Falcone, A. D., Barthelmy, S. D., Burrows, D. N., et al. 2006, GCN\,4966
\bibitem[Fan et al.(2004)]{Fan04} Fan, Y.~Z., Wei, D.~M., \& Wang, C.~F.\ 2004, A\&A, 424, 477 
\bibitem[Fenimore et al.(1993)]{Fenimore93} Fenimore, E. E., Epstein, R. I., \& Ho, C. 1993, A\&AS, 97, 59 
\bibitem[Frail et al.(1997)] {Frail97} Frail, D. A., Kulkarni, S. R., Nicastro, L., Feroci, M., \& Taylor, G. B. 1997, Nature, 389, 261
\bibitem[Gehrels et al.(2004)]{Neil04} Gehrels, N., Chincarini, G., Giommi, P., et al. 2004, ApJ, 611, 1005 
\bibitem[Golenetskii et al.(2006)]{Golen06} Golenetskii, S., Aptekar, R., Mazets, E., et al. 2006, GCN\,4989
\bibitem[Granot \& K\"onigl(2003)]{GrKo03} Granot, J., \& K\"onigl, A. 2003, ApJ, 594, 83
\bibitem[Guetta et al.(2001)]{Guetta01} Guetta, D., Spada, M., \& Waxman, E. 2001, ApJ, 557, 399
\bibitem[Jin \& Fan(2007)]{Jin07} Jin, Z.-P., \& Fan, Y.-Z. 2007, MNRAS, 378, 1043
\bibitem[Lazzati et al.(2004)]{Laz04} Lazzati, D., Covino, S., Gorosabel, J., et al., 2004, A\&A, 422, 121
\bibitem[Ledoux et al.(2006)]{Led06} Ledoux, C., Vreeswijk, P., Smette, A., Jaunsen, A., \& Kaufer, A. 2006, GCN\,5237
\bibitem[Lithwick \& Sari(2001)]{Lith01} Lithwick, Y., \& Sari, R. 2001, ApJ, 555, 540 
\bibitem[Malesani et al.(2007)]{Mal07} Malesani, D., Molinari, E., Vergani, S.D., et al., 2007, astro-ph/0706.1772
\bibitem[M\'esz\'aros \& Rees(1997)]{MesRees97} M\'esz\'aros, P., \& Rees, M.~J. 1997, ApJ, 476, 232
\bibitem[M\'esz\'aros(2006)]{Meszaros} M\'esz\'aros, P. 2006, Rep. on Progr. in Phys. 69, 2259
\bibitem[Molinari et al.(2007)]{Mol07} Molinari, E., Vergani, S.D., Malesani, D., et al. 2007, A\&A, 469, 13
\bibitem[Mundell et al.(2007)]{Mun07} Mundell, C.G., Steele, I.A., Smith, R.J., et al. 2007, Science, 315, 1822
\bibitem[Panaitescu \& Kumar(2000)]{PK00} Panaitescu, A., \& Kumar, P. 2000, ApJ, 543, 66
\bibitem[Piran(2000)]{Piran00} Piran, T. 2000, Phys. Rep., 333, 529
\bibitem[Rees \& M\'esz\'aros(1994)]{Rees94} Rees, M. J., \& M\'esz\'aros, P. 1994, ApJ, 430, L93 
\bibitem[Roming et al.(2006)]{Rom06} Roming, P.W.A., Schady, P., Fox, D. B., et al. 2006, ApJ, 652, 1416
\bibitem[Ruderman(1975)]{Ruderman75} Ruderman, M. 1975, NYASA, 262, 164
\bibitem[Sagiv et al.(2004)]{Sag04} Sagiv, A., \& Waxman, E., Loeb, A. 2004, ApJ, 615, 366
\bibitem[Sari \& Piran(1999)]{SP99} Sari, R., \& Piran, T. 1999, ApJ, 520, 641
\bibitem[Soderberg \& Ramirez-Ruiz(2002)]{Alicia02} Soderberg, A.~M., \& Ramirez-Ruiz, E. 2002, MNRAS, 330, L24
\bibitem[Taylor et al. (2005)] {Taylor05} Taylor, G. B., Momjian, E., Pihlstrom, Y., Ghosh, T., \& Salter, C. 2005, ApJ, 622, 986 
\bibitem[Tueller et al.(2006)]{Tuel06} Tueller, J., Barbier, L., Barthelmy, S., et al. 2006, GCN\,5242
\bibitem[Vreeswijk \& Jaunsen (2006)]{Vrees06} Vreeswijk, P., \& Jaunsen, A. 2006, GCN\,4974
\bibitem[Zerbi et al.(2001)]{Zerb01} Zerbi, F. M., Chincarini, G., Ghisellini, G., et al. 2001, AN, 322, 275
\bibitem[Zhang \& Kobayashi(2005)]{ZhangKob05} Zhang, B., \& Kobayashi, S. 2005, ApJ, 628, 315 
\bibitem[Zhang et al.(2006)]{zhang06} Zhang, B., Fan, Y.~Z., Dyks, J., et al. 2006, ApJ, 642, 354.
\bibitem[Ziaeepour et al.(2006)]{Ziaee06} Ziaeepour, H. Z., Barthelmy, S. D., Gehrels, N., et al. 2006, GCN\,5233
\end{thebibliography}
\end{document}